\documentstyle[prl,aps,preprint,tighten]{revtex}
\begin{document}
\draft

       \title{ On the low temperature properties and specific anisotropy of
pure anisotropically paired superconductors}
\author{Yu.S.Barash, and A.A.Svidzinsky \\  }
\address {I.E. Tamm Department of Theoretical Physics, P.N. Lebedev Physics
Institute \\
     Leninsky Prospect 53, Moscow 117924, Russia \\ }
\maketitle
\begin{abstract}
Dependences of low temperature behavior and anisotropy of various  physical
quantities for pure unconventional superconductors upon a particular form of
momentum direction dependence for the superconducting order parameter
(within the framework of the same symmetry type of superconducting pairing)
are considered. A special attention is drawn to the
possibility of different multiplicities of the nodes of the order parameter
under their fixed positions on the Fermi surface, which are governed by
symmetry.
The problem of an unambiguous identification of a type of superconducting
pairing on the basis of corresponding experimental results is discussed.
Quasiparticle density of states at low energy for both homogeneous and mixed
states, the low temperature dependences of the specific heat, penetration depth
and thermal conductivity, the I-V curves of SS and NS tunnel junctions at low
voltages are examined. A specific anisotropy of the boundary conditions for
unconventional superconducting order parameter near $T_c$ for the case of
specular reflection from the boundary is also investigated.
 \end{abstract}
 \narrowtext
\section{Introduction}
The possibility of unconventional types of superconducting pairing in a number
of heavy-fermion and high-$T_c$ compounds, which follows from many experimental
observations, results in much theoretical attention to a study of various
physical properties of unconventional superconductors. If an unconventional
superconducting order parameter $\hat\Delta(\bbox{p})$ vanishes at points or
lines on the Fermi surface, low-energy quasiparticle excitations in the
vicinities of these nodes may exist and are known to give rise to power laws for
low temperature dependences of the specific heat $C(T)$, the thermal
conductivity $\kappa(T)$ and some other quantities to be measured. Indices for
these power laws depend not only on the dimension of nodes (that is whether
there are points or lines of nodes) or, for example, on the strength of the
impurity scattering. They differ also for the order parameters,
which have zeroes of different orders at the nodes. Since always, even within
the same symmetry type of pairing, there are basis functions which differ from
each other by the multiplicities of nodes situated at the same lines and points
on the Fermi surface, different indices may correspond to the same type of
pairing.
It seems to be quite important as different experimental consequences for
superconductors with the same type of pairing become possible for different
particular basis functions for the order parameter. For instance, let's
consider $(p_x^2-p_y^2)$-type of pairing in a tetragonal superconductor. For the
basis function ($p_x^2-p_y^2$) and a spherical Fermi surface one gets at the
Fermi surface $\Delta=\Delta_0\sin^2\theta\cos2\varphi$. Near lines of nodes
$\varphi_{1,2}=\pm\pi/4$ and far from poles of the sphere this order parameter
is proportional to $(\varphi-\varphi_{1,2})$ and near the intersections of lines
at the poles one has $\Delta\propto (\varphi-\varphi_{1,2})\theta^2$. Other
basis functions for the same representation of a symmetry group $D_{4h}$ and a
spherical Fermi surface, for example,
$\Delta=\Delta_0\sin^{2m}\theta\cos^{2n+1}2\varphi$ \ ($n, m=0,1,2\ldots$) have
obviously different orders of
zeroes at the nodes and, consequently, one obtains different low-temperature
behaviors of various physical quantities  for these basis functions. The same
arguments are applicable, of course, to the case of an arbitrary shape of the
Fermi surface, for instance, to the cylindrical Fermi surface, when instead of
the simplest representative for the order parameter
$\Delta=\Delta_0\cos2\varphi$ for the case of $(p_x^2-p_y^2)$-type of
pairing, one can consider more complicated examples like
$\Delta=\Delta_0\cos^{2n+1}2\varphi$. Due to this
fact an identification of the type of superconducting pairing (and frequently
even a distinction between the contributions from lines and points of nodes) on
the basis of experimental data for low temperature behavior of various
quantities turns out to be ambiguous or at least more complicated as it is
usually supposed. So, it is of importance to investigate the influence of the
order parameter behavior near the nodes on some properties of unconventional
superconductors.

Having in mind this circumstance we calculate below the quasiparticle density of
states at low energy and low temperature dependences of the specific heat ( for
both homogeneous and mixed states ), the penetration depth, the thermal
conductivity and the I-V curves for
SS and NS tunnel junctions at low voltages, permitting the order parameter to
have zeroes of different orders at the nodes on a Fermi surface. We consider
also within this context an anisotropy of boundary conditions for an
unconventional superconducting order parameter near $T_c$ for the case of
specular reflection. Our results allow to carry out a qualitative and simple
analysis of some experimental data, taking account of the possibilities
mentioned above. We believe that such a consideration, which permits to
discuss qualitatively experimental consequences for a wide set of the basis
functions, is needed along with the detailed quantitative (usually numerical)
calculations of the effects for several currently preferable  representatives
from the set for the given compound. In particular, we discuss recent
experimental data on the low temperature behavior of thermal conductivity for
the heavy fermion superconductor $UPt_3$\cite{lussier1,lussier2}.

\section{Density of states and specific heat}

If a superconducting order parameter doesn't vanish anywhere on the Fermi
surface (as it is in the particular case of an isotropic s-wave superconductor),
the density of states $N(E)$ is equal to zero at low energies $E<\Delta_{min}$
and the electronic specific heat $C(T)$ falls off exponentially with the
decrease of temperature
$C\propto e^{-\Delta_{min}/T}$, since only a small number of quasiparticles are
activated above the gap at $T\ll\Delta_{min}$.
For an anisotropic pairing with nodes of the order parameter on the Fermi
surface the low energy density of states is known to be nonzero. The
contributions to $N(E)$ at low energy  come from quite narrow vicinities of the
nodes, they dominates also in $C(T)$ at low temperatures. Besides, in the case
of unconventional
superconductor the scattering on nonmagnetic impurities causes a substantial
change in the low-energy density of states for various scattering strength and
even at quite small impurity concentrations\cite{uedarice,hir2,muz1}. This
underlies the low temperature behavior for various physical quantities
including the specific heat. Below we consider only a pure unconventional
superconductor which is relevant even to the case of resonance scattering at
low impurity density, at least for the energies
$\omega_c\lesssim E\ll T_c$ above a critical energy
$\omega_c=(\Gamma\Delta_0)^{1/2}$ (where $\Gamma$ is the normal state scattering
rate in the unitarity limit).

For a unitary gap matrix $\hat\Delta(\bbox{p})$ and a spherical Fermi surface
the quasiparticle density of states in the case of pure superconductor is given
by
\begin{equation}
N(E)=2\sum_{\bbox{p}}\delta(E-E_{\bbox{p}})=N_F\int_{|\Delta(\bbox{p})|<E}
\frac{\displaystyle d\Omega}{\displaystyle 4\pi}
\frac{\displaystyle E}{\displaystyle \sqrt{E^2-|\Delta(\bbox{p})|^2}} \quad ,
\label{density}
\end{equation}
where the integration is carried out over the momentum directions and $N_F$ is
a density of states at the Fermi surface for normal metal.

The electronic part of the specific heat may be written as
$$
C(T)=2\int_0^{\infty}dE\left\{N(E)
\frac{\displaystyle E^2\exp (E/T)}{\displaystyle T^2\left(\exp (E/T)+1\right)^2}
+T\frac{\displaystyle dN(E)}{\displaystyle dT}\biggl [\ln(1+\exp (-E/T))+
\right.
$$
\begin{equation}
\left.
+\frac{\displaystyle E}{\displaystyle T(1+\exp (E/T))}\biggr ]\right\} .
\label{specific}
\end{equation}

One can easily estimate low energy and low temperature behavior of quantities
$N(E)$ and $C(T)$ setting the particular form of the gap function only in the
vicinities of nodes. For example, if the order parameter
in the vicinities of lines of nodes takes the form ($n>0$)
\begin{equation}
|\Delta(\bbox{p})|=\Delta_0|\varphi-\varphi_0|^n ,
\label{line1}
\end{equation}
one easily gets the main contributions to $N(E)$, $C(T)$:
\begin{equation}
\frac{\displaystyle N(E)}{\displaystyle N_F}=4h(n)\left(E/\Delta_0\right)^{1/n},
\end{equation}
\begin{equation}
C(T)=N_F4w(n)T\left(T/\Delta_0\right)^{1/n}.
\label{C1}
\end{equation}
Here the functions $h(x)$, $w(x)$ are given by
\begin{equation}
h(x)=\frac{\displaystyle \Gamma(1/2x)}{\displaystyle
8\sqrt{\pi}x\Gamma((x+1)/2)} ,
\label{h}
\end{equation}
\begin{equation}
w(x)=2h(x)\left(2+\frac{\displaystyle 1}{\displaystyle x}\right)
\left(1-\frac{\displaystyle 1}{\displaystyle 2^{1+1/x}}\right)\Gamma(2+1/x)
\zeta(2+1/x),
\label{w}
\end{equation}
where $\Gamma(x)$, $\zeta(x)$ are the Gamma function and the Riemann Zeta
function respectively.

Analogously, if near the lines of nodes one has ($n>0$)
\begin{equation}
|\Delta(\bbox{p})|=\Delta_0|\theta-\theta_0|^n ,
\label{line2}
\end{equation}
the corresponding contributions to the density of states and specific heat are
\begin{equation}
\frac{\displaystyle N(E)}{\displaystyle N_F}=\sin\theta_04\pi h(n)
\left(E/\Delta_0\right)^{1/n},
\end{equation}
\begin{equation}
C(T)=N_F4\pi\sin\theta_0w(n)T\left(T/\Delta_0\right)^{1/n}.
\label{C2}
\end{equation}

There are obviously other basis functions belonging to the same representation
of the symmetry group $D_{4h}$. So, let the order parameter near the nodes on
the Fermi sphere within the small angle regions  $|\phi-\phi_0|\le b$, \
$\theta^2\le a$ ($a,b\ll 1$) be

\begin{equation}
|\Delta(\bbox{p})|=\Delta_0|\varphi-\varphi_0|^m\theta^{2n} ,
\label{1}
\end{equation}

where $m,n\ge 0$, $m+n>0$. Then one obtains from Eqs.(\ref{density}),
(\ref{specific}):
\begin{equation}
\frac{\displaystyle N(E)}{\displaystyle N_F}=
\left\{
\begin{array}{ll}
\left(E/\Delta_0\right)^{1/m}\frac{
\displaystyle a^{1-\frac{n}{m}}}{(1-\frac{n}{m})}h(m), &m>n\\
\\
\left(E/\Delta_0\right)^{1/n}\frac{
\displaystyle 1}{\displaystyle n}\left(\ln(\Delta_0/E)+c\right)h(n) ,\ \ \ \  &m=n\\
\\
\left(E/\Delta_0\right)^{1/n}\frac{
\displaystyle b^{1-\frac{m}{n}}}{(1-\frac{m}{n})}h(n) , &n>m,\\
\end{array}
\right.
\label{den1}
\end{equation}
\begin{equation}
C(T)=
\left\{
\begin{array}{ll}
N_FT\left(T/\Delta_0\right)^{1/m}\frac{
\displaystyle a^{1-\frac{n}{m}}}{(1-\frac{n}{m})}w(m), &m>n\\
\\
N_FT\left(T/\Delta_0\right)^{1/n}\frac{
\displaystyle 1}{\displaystyle n}\left(\ln(\Delta_0/T)+c\right)w(n) ,\ \ \ &m=n\\
\\
N_FT\left(T/\Delta_0\right)^{1/n}\frac{
\displaystyle b^{1-\frac{m}{n}}}{(1-\frac{m}{n})}w(n) , &n>m.\\
\end{array}
\right.
\label{heatcap1}
\end{equation}

A role of logarithmic terms in Eqs.(\ref{den1}), (\ref{heatcap1}) depends not
only upon parameters
$E/\Delta_0$, $T/\Delta_0$ but also on the magnitude of a constant $c$,
which is formed by the contributions from the angular vicinity of the
line of nodes, but not only very close to the pole. This constant may be
calculated by making use of particular form of the basis function all over the
Fermi sphere. For
instance, in the case of $(p_x^2-p_y^2)$-type of pairing with
$\Delta=\Delta_0\sin^2\theta\cos2\varphi$ one gets from (\ref{density})
\begin{equation}
\frac{\displaystyle N(E)}{\displaystyle N_F}=\left(A+\ln (\Delta_0/E)
\right)
\frac{\displaystyle E}{\displaystyle 2\Delta_0} ,
\label{d}
\end{equation}
where
\begin{equation}
A=\frac{\displaystyle 2}{\displaystyle \pi}\int_0^1dk\left[K(k)-
\frac{\displaystyle \ln k}{\displaystyle k}
\left(\frac{\displaystyle E(k)}{\displaystyle 1-k^2}-
K(k)\right)\right]\simeq1.38 .
\label{constant}
\end{equation}
Here $K(k)$ and $E(k)$ are the complete elliptic integrals of first and second
type.

So, both terms in (\ref{d}) may be of the same order and, as a rule, they both
should be taken into account. Eq.(\ref{d}) is in accordance with the second
relation in (\ref{den1}), since the order parameter $\Delta$ near the nodes
$\theta_0=0$, $\varphi_{1,2}=\pm \pi/4$ takes the same form as in (\ref{1})
with $n=m=1$, and the logarithmic term in (\ref{d}) contains respective
contributions from all four nodes (that is from two lines of nodes at two
poles of a Fermi sphere ). As in this example, the behavior of the order
parameter of the form (\ref{1}) usually corresponds to the intersection of two
lines of nodes at the pole (each being of the form
$|\Delta(\bbox{p})|=\Delta_0|\varphi-\varphi_0|^m $ near the line and far from the
pole). One can easily seen that under the condition $n\geq m$ the part coming
from the point of intersection dominates or is at least of the same order as
the total contribution to the quantities.
Note, that in the particular case $m=0$  Eqs.(\ref{den1}), (\ref{heatcap1})
at $n>m$ correspond to the contributions from the point node which is described
by Eq.(\ref{line2}) for $\theta_0=0$ and the substitution $n\rightarrow 2n$.

At last, for the order parameter which may be represented near the point node as
follows
\begin{equation}
|\Delta(\bbox{p})|=|\Delta_{01}(\varphi-\varphi_0)^m+\Delta_{02}(\theta-
\theta_0)^n| ,  \ \ \ \ \ \ m,n\ge 0, \ \ \ m+n>0,
\label{point}
\end{equation}
one gets
\begin{equation}
\frac{\displaystyle N(E)}{\displaystyle N_F}\propto \sin\theta_0E^{1/n+1/m},
\ \ \ \
C(T)\propto\sin\theta_0T^{1+1/n+1/m}.
\label{C3}
\end{equation}

Summing up the contributions from all vicinities of the nodes of the order
parameter, one finds total low energy density of states and low-temperature
specific heat.

In discussing lines of nodes of the order parameter one usually considers
zeroes of the first order at the nodes which correspond to relation
(\ref{line1}) or (\ref{line2}) with the particular value $n=1$.
Then, according to the Eqs.(\ref{C1}), (\ref{C2}) one gets a quadratic
low-temperature behavior for the heat capacity $C(T)\propto T^2$. Analogously,
for the points of nodes it is frequently supposed $n=m=1$ in (\ref{point}) and
then one obtains from (\ref{C3}) the well-known result $C(T)\propto T^3$.
As it is seen from the more general relations (\ref{C1}), (\ref{C2}),
(\ref{C3}), the possibility for distinguishing between the contributions
from points and lines of nodes becomes substantially more restricted, and
indices of the power laws are allowed to be fractional, even for integer
values of $n, m$. For instance, for an experimental determination whether there
is a contribution from a cubic point of nodes ($m=n=3$, when corresponding term
in the specific heat is $C_{cp}\propto T^{5/3}$), or from a first order line of
nodes (when one has $C_{fl}\propto T^{2}$), quite a high accuracy is needed as
compared to the distinguishing between the heat capacity dependences
$C_{fp}\propto T^{3}$ and $C_{fl}\propto T^{2}$.
Note that a contribution to the specific heat from zeroes of
second order at points (quadratic points) is proportional to $T^2$ precisely as
from the first order line of nodes.
In the case of fractional values of
$n$, $m$ as well as for the dependences like $|\theta -\theta_0|$ the question
about the physical origin for the nonanalitical behavior of the order parameter
near the nodes appears.

\section{Low energy density of states and low-temperature specific heat for the
mixed state}

One of the most important qualitative features of an unconventional
superconductor in a mixed state is that due to the existence of a supercurrent
induced by a magnetic field, the quasiparticle states with negative energy
(which are obviously occupied) appear for momentum directions within the
vicinity of nodes of the order parameter, as it follows from the relation
$E(\bbox{p})=\sqrt{\xi(\bbox{p})^2+|\Delta(\bbox{p})|^2}+
\bbox{p}\bbox{v}_s$\cite{saulyip}.
The density of states for such a superconductor changes substantially
at sufficiently low energies. In particular, the density of states
at the Fermi surface differs from zero under these conditions and after spatial
averaging over the mixed state is proportional in the simplest case to
$\sqrt{B}$, resulting in the characteristic magnetic
field dependent term in the specific heat $C\propto T\sqrt{B}$\cite{vol}.
Recently the term of such kind, which is specific for unconventional
superconductors, has been observed experimentally for $YBCO$\cite{mol,mol2}. From
these measurements a quite strong upper limit on the minimum gap value follows
if an anisotropic $s$-wave superconductivity is assumed. We examine below
the influence of the particular forms of basis functions, which belong to the
same symmetry type of pairing, on the magnetic field dependence of the density
of states and heat capacity (under the conditions similar to those considered
in\cite{vol}). It will be shown that an accurate experimental determination of
the index value in this power law dependence permits to describe the behavior of
the order parameter in the vicinity of nodes more carefully. The spatial
distribution of the density of states around the vortex is also considered.

Let a strong type two pure superconductor with a cylindrical Fermi surface be
at low temperature in a mixed state under the applied magnetic field, which
satisfies the condition $H_{c1}\lesssim B\ll H_{c2}$ and is directed along the
principal cylindrical axis. For this inhomogeneous state
the local density of states must obviously manifest a spatial dependence and,
generally speaking, is formed by both a superconducting region outside the
vortex cores and the quasiparticle states localized inside the cores. Since the
total volume occupied by the vortex cores is proportional to the magnetic field,
the contribution to the density of states from the states localized within the
vortex cores is a linear function of $B$ and the corresponding contribution to
the specific heat $\propto TB/H_{c2}$\cite{fetter}.
The quasiparticles at distances $r\gg \xi_0$ from the vortex core may be
considered quasiclassically as having the energy
\begin{equation}
E(\bbox{p,r})=\sqrt{\xi(\bbox{p})^2+|\Delta(\bbox{p})|^2}+
\bbox{p}\bbox{v}_s(\bbox{r}) ,
\label{le}
\end{equation}
which depends upon the distance from the vortex core along with the
superfluid velocity (in the simplest case of circular supercurrents one has
$\bbox{v}_s=\bbox{e}_{\varphi}K_1(r/\lambda)/2m_e\lambda $). Then the
corresponding density of states may be written as follows
\begin{equation}
N(E,\bbox{r})=2\int\frac{\displaystyle d^3p}{\displaystyle (2\pi)^3}
\delta\left(\sqrt{\xi^2(\bbox{p})+|\Delta(\bbox{p})|^2} +
\bbox{p}\bbox{v}_s(\bbox{r})-E\right).
\label{d1}
\end{equation}

From the estimate
$|\bbox{p}\bbox{v}_s|\lesssim p_F/m_er\ll v_F/\xi_0\sim \Delta_0$
which is valid at distances $r\gg\xi_0$ it follows, that a nonzero contribution
to the density of states for the zero quasiparticle energy appears only for
unconventional superconducting pairing for momentum directions within the narrow
vicinities of nodes of the order parameter $\Delta(\bbox{p})$ ( $\Delta_0$
denotes throughout this paper the maximum value of the order parameter for the
homogeneous superconducting state at low temperature and for any type of
pairing). Due to this fact
in considering the density of states at quite low energy one can describe the
contributions from each line or point of nodes separately.

The location of a line of nodes of the order parameter, which is oriented
parallel to the principal axis of the cylindrical Fermi surface, may be
described by a constant polar angle $\varphi_l$ of the cylindrical coordinate
system, corresponding to the momentum direction $\bbox{p}_{F,\perp}(l)$ to the
line. In integrating in (\ref{d1}) one can put  with a good accuracy
$\bbox{p}\bbox{v}_s=\bbox{p}_{F,\perp}(l)\bbox{v}_s$ due to a quite small
parameter $v_s/v_F$. Suppose that the quasiparticle energy $\xi(\bbox{p})$
for the normal metal and the order parameter $\Delta(\bbox{p})$ near nodes
depend upon the different momentum components -- the magnitude of the momentum
component $p_r$ and the direction of the momentum ${\bbox{p}}_{F,\perp}$
(which is perpendicular to the principal cylindrical axis) correspondingly.
Then one obtains after integration over the energy $\xi(\bbox{p_r})$ ($d\xi=
v_{F}dp_r$) the following contribution from the line of nodes
\begin{equation}
N(E,\bbox{r})=\frac{\displaystyle N_F}{\displaystyle \pi}
\int_{\Omega_l} d\varphi \frac{\displaystyle |E-\bbox{p}_{F,\perp}(l)\bbox{v}_s|
\Theta(E-\bbox{p}_{F,\perp}(l)\bbox{v}_s)}{\displaystyle \sqrt{(E-
\bbox{p}_{F,\perp}(l)\bbox{v}_s)^2-|\Delta (\varphi)|^2}}.
\label{d2}
\end{equation}
Here $\Theta(x)$ is the step-like function, $N_F$ is the density of states at the
cylindrical Fermi surface for the normal metal and $\Omega_l$ is the narrow
angular region in the vicinity of the line of nodes, where the expression under
the square root sign in (\ref{d2}) is positive at quite low energies in
question.

If in the close vicinity of the nodes the order parameter takes the form
\begin{equation}
|\Delta(\varphi)|=\Delta_0|\varphi-
\varphi_{l}|^n, \ \ \ \ n>0 ,
\label{op}
\end{equation}
one gets from (\ref{d2}) after the integration over $\varphi$
\begin{equation}
N(E,\bbox{r})=\frac{\displaystyle N_F\Gamma\left(\frac{1}{2n}
\right)}{\displaystyle n\sqrt{\pi}\Gamma\left(\frac{1}{2}+\frac{1}{2n}\right)}
\left(\frac{\displaystyle E-\bbox{p}_{F,\perp}(l)\bbox{v}_s(\bbox{r})}{
\displaystyle\Delta_0}\right)^{1/n}\Theta(E-\bbox{p}_{F,\perp}(l)
\bbox{v}_s(\bbox{r})) .
\label{d3}
\end{equation}

According to this relation, the narrow angular region in the momentum space
situated near the line of nodes on the Fermi surface, contributes to the local
quasiclassical density of states at sufficiently low energy for quite a wide
region of orientations in the coordinate space. For instance, at zero energy
the density of states formed by the line of nodes is nonzero for the whole
spatial region defined by the condition
$\bbox{p}_{F,\perp}(l)\bbox{v}_s(\bbox{r})<0$, though
it naturally falls off with the increasing distance from the vortex core (along
with the supecurrents induced by the magnetic field). So, in the presence
of several lines of nodes the contributions are spatially superimposed and due
to this fact the locations  of a maximum value of the total density of states
in the coordinate space (at the  given distance $r$ from the vortex core) may
substantially differ from the locations of corresponding maxima of the partial
contributions from each separate line of nodes.

In the case of $(p_{x}^2-p_{y}^2)$-pairing for the tetragonal superconductor
with cylindrical Fermi surface, one can consider four similar lines of nodes
for the superconducting order parameter, which are situated at the angles
$\varphi_l=\pi/4+(l-1)\pi/2$ \ ($l=1,2,3,4$). Let the angle $\phi$ specifies
the relative orientation of the vector $\bbox{r}$ and the crystalline
axis $x$, and $v_s(\bbox{r})=v_s(r)\bbox{e}_{\varphi}$. Then the sum of all four
terms of the form (\ref{d3}) results in the following spatial dependent density
of states
$$
N(E,r,\phi)= \qquad \qquad \qquad \qquad \qquad \qquad \qquad \qquad \qquad \qquad \qquad \qquad \qquad \qquad \qquad \qquad \qquad
$$
\begin{equation}
=\frac{\displaystyle N_F\Gamma\left( \frac{1}{2n}
\right)}{\displaystyle n\sqrt{\pi}\Gamma\left(\frac{1}{2}+
\frac{1}{2n}\right)}\left( \frac{\displaystyle v_FK_1(r/\lambda)}{\displaystyle
2\lambda \Delta_0 } \right) ^{1/n} \sum_{l}\left( \tilde E+\sin (\phi+
\varphi_l)\right )^{1/n}\Theta\left(\tilde E+\sin (\phi+\varphi_l)
\right) ,
\label{d4}
\end{equation}
where the dimensionless quantity $\tilde E= 2\lambda E/v_F K_1(r/\lambda)$ is
introduced.

Since at large distanses from the vortex core  the anisotropy
of $N$ is small, we
consider further only the distances $r\ll \lambda$. At the zero energy one gets from here
$N\propto N_F(\xi_0/r)^{1/n}\left(|\cos (\phi +\pi/4)|^{1/n}+|\sin (\phi +
\pi/4)|^{1/n}\right)$ and in the particular
case $n=1$ the spatial angular dependence of $N$ is simply
$N\propto |\cos\phi |$ for $-\pi /4<\phi +\pi m<\pi /4$ and
$N\propto |\sin\phi |$ for $\pi /4<\phi +\pi m<3\pi /4$ ($m=0,1$).
Hence, the maximum value of the density of states at given $r$ lies at
$\phi= 0,\pm \pi/2,\pi $  and
minimum - at $\phi=\pm \pi/4, \pm 3\pi/4$. At the same time the maximum of
the separated contribution from each line of nodes lies  in the
coordinate space at the direction, where the superfluid velocity is opposed to
the momentum direction to the corresponding
line on the Fermi surface (i.e. all four these maxima lie in the directions
$\phi=\pm \pi/4, \pm 3\pi/4$). Furthermore, in the particular case
$n=1$ and for $\tilde E>1$ (e.g. at the distances $r>(T_c/E)\xi_0$) all
spatial dependence of the density of states (both from the distance and the
direction) disappears and one gets from (\ref{d4})
$N\sim (E/\Delta_0)N_F$. It follows from here that the greater the energy
the smaller the spatial region where the anisotropy and inhomogeneity of the
density of states manifest themselves.
So, for $E\sim T_c$ one has the anisotropy only within the region
$r<\xi_0$, where the local description based on the relation (\ref{le}) for the
quasiparticle energy, can't be applied and more general approach is
needed\cite{rem1}. Therefore, within the framework of the approach used above
one may consider only the energies $E\ll T_c$. It is worth noting that in
contrast to the particular case $n=1$, for $n\ne 1$ the angular and distance
dependences of the density of states at large distances don't disappear
completely though become quite weak.
The spatial dependence  of the density of states (\ref{d4}), including its
fourfold symmetry, is accessible for studying to scanning tunneling microscopy
experiments, which now have a high spatial resolution as a valuable complement
to the high energy resolution of tunneling spectroscopy (see, for example,
\cite{hhess,fischer}).

As the density of states at zero energy is finite, then one gets from
(\ref{specific}) the conventional (linear in $T$) expression for the
low-temperature specific heat $C(T)= \pi^2T\tilde N(0)/3$. Due to the
inhomogeneity of the state the spatial average of the density of states
$\tilde N(0)$ is presented here. For the magnetic fields within the
interval $H_{c1}\ll B\ll H_{c2}$ one substitutes $\lambda/r$ instead of
$K_1(r/\lambda)$ in (\ref{d3}), (\ref{d4}) and then the spatial averaging over
the region $\xi_0\ll r\ll R$ (where $R\sim\xi_0 (H_{c2}/B)^{1/2}$ is the distance
between vortices) gives rise to the following estimate
\begin{equation}
\tilde N(0)=A_nN_F\left(\frac{\displaystyle v_F}{\displaystyle \Delta_0\xi_0}
\right)^{1/n}\left(\frac{\displaystyle B}{\displaystyle H_{c2}}\right)^{1/2n} .
\label{ad}
\end{equation}
Here $A_n$ depends only upon the index $n$. Note, that in forming $\tilde N(0)$
just the distances $r\gg \xi_0$ are found to be important for $n>1/2$.
For these values of $n$ the contribution to the specific heat
$\propto T(B/H_{c2})^{1/2n}$ is more essential, than the linear magnetic field
dependence, coming from the interiors of the vortex cores.

According to (\ref{ad}), the index $1/2n$ of the power law for the magnetic
field dependence of the low-temperature specific heat for the unconventional
superconductor in the mixed state characterizes not only the existence of
nodes of the order parameter on the Fermi surface themselves, but also the
behavior of the order parameter in the vicinities of the nodes, for example,
of a form (\ref{op}).

\section{Penetration depth}

It is well known that deviation of the penetration depth at low temperatures
from its zero temperature value is exponentially small in the case of a
superconductor with finite gap and manifests a power law temperature
dependence for a superconductor with nodes of the order parameter on the
Fermi surface\cite{gross,choimuz,procar,hirgol,xu}. We are interested here
in the indices of those power laws for temperatures $(\Gamma\Delta_0)^{1/2}
\lesssim T\ll \Delta_0$ in the case of pure superconductors with
unitary order parameters having various multiplicities of the nodes.
As a starting point one can use the following expression for the eigenvalues
of the penetration depth tensor:
\begin{equation}
\frac{\displaystyle 1}{\displaystyle\lambda_{i}^2}=
\frac{\displaystyle 4\pi e^2}{\displaystyle c^2} \int
\frac{\displaystyle d^2 S}{\displaystyle (2\pi)^2v_F} v_{F,i}^2 T \sum_m
\frac{\displaystyle |\Delta(\hat{\bbox{p}})|^2}{\displaystyle\left(\left(2m+
1\right)^2\pi^2T^2+|\Delta(\hat{\bbox{p}})|^2\right)^{3/2}}  ,
\label{l1}
\end{equation}
where index $i$ denotes the direction of the supercurrent and
$ v_{F,i} $ is the $ i $-component of the quasiparticle velocity at the
Fermi surface.
Applying the Poisson formula to the sum in (\ref{l1}),
we get for the deviation $\lambda_i (T)$ from  $\lambda_i (0)$  at
$ T \ll T_{c}$ :

\begin{equation}
\frac{\displaystyle \delta\lambda_{i}(T)}{\displaystyle\lambda_{i}(0)}=
- \ \frac{\displaystyle \int\frac{\displaystyle d^2 S}{\displaystyle v_{F}}
v_{F,i}^2 \sum_{m=1,2,\ldots }(-1)^m mK_1\left(m\frac{\displaystyle
|\Delta(\hat{\bbox{p}})|}{ \displaystyle T}\right) \frac{\displaystyle
|\Delta(\hat{\bbox{p}})| }{\displaystyle  T}} {\displaystyle \int\frac{\displaystyle
d^2 S}{\displaystyle v_{F}} v_{F,i}^2  } ,
\label{l2}
\end{equation}
where $K_1(x)$ is the modified Bessel function.

For $T\ll \Delta_0 $ the main contribution to the integral over the Fermi
surface in the numerator comes from narrow regions near nodes of
$ \Delta(\hat{\bbox{p}}) $.
As in the previous section we consider a tetragonal superconductor with
cylindrical Fermi surface and the order parameter having four lines of nodes.
Then it follows from (\ref{l2}) for the in-plane penetration depth

\begin{equation}
\frac{\displaystyle \delta\lambda_{\Vert}(T)}{\displaystyle\lambda_{\Vert}(0)}=
-\int_{0}^{2\pi}\frac{\displaystyle d\varphi}{\displaystyle
2\pi}\sum_{m=1,2,\ldots }(-1)^m mK_1\left(m\frac{\displaystyle
|\Delta(\varphi)|}{ \displaystyle T}\right) \frac{\displaystyle
|\Delta(\varphi)| }{\displaystyle  T} ,
\label{l3}
\end{equation}
where the integration is carried out over the angle in the basal plane.

Assuming relation (\ref{op}) for the order parameter in the vicinities of lines
of nodes we get for the leading correction to the penetration depth at low
temperatures

\begin{equation}
\frac{\displaystyle \delta\lambda_{\Vert}(T)}{\displaystyle\lambda_{\Vert}(0)}=
\Omega (n) \left( \frac{\displaystyle T }{\displaystyle \Delta_0 }\right)^{
1/n } ,
\label{l4}
\end{equation}

where
$$
   \Omega (n)=-\frac{\displaystyle 2}{\displaystyle \pi n}2^{1/n}
\Gamma\left( \frac{\displaystyle 2n+1 }{\displaystyle 2n }\right)
\Gamma\left( \frac{\displaystyle 1 }{\displaystyle 2n }\right)
\sum_{m=1,2,\ldots}
\frac{\displaystyle (-1)^m }{\displaystyle m^{1/n} }.
$$

In the particular case $n=1$ one obtains from here  $\delta\lambda_{\Vert}(T)/
\lambda_{\Vert}(0)=2\ln 2
\left (T/\Delta_{0}\right)$.

\section{Thermal conductivity}

The investigation of the thermal conductivity in a superconducting state is an
important experimental probe of the order parameter structure.

Thermal conductivity for an isotropic s-wave superconductor is known to
fall off exponentially at low temperatures due to the presence of a
finite superconducting gap for all directions on the Fermi surface.
For unconventional superconductors a temperature dependence of the electron
thermal conductivity
at low temperatures $T\ll \Delta_0$  is described by the power laws and proves
to be associated with several factors. Besides the order parameter
structure, it is substantially  influenced by the
strength of impurity scattering. For the thermal conduction components directed
to the nodes of the order parameter one usually obtains in the case of weak
scattering  processes (Born approximation) a linear
temperature dependence, as in the normal state. Other components vanish with the
temperature according to the higher power laws. In the case of resonance
impurity scattering the situation is more complicated and the region of low
temperatures $T\ll \Delta_0$ is naturally devided into two parts. For
temperatures $T\lesssim \omega_c=(\Gamma\Delta_0)^{1/2}$ the existence of the
quasiparticle states bound to impurities is of importance. The correspondent
thermal conductivity at those temperatures has also linear temperature
dependence albeit with reduced density of states. By contrast, for the
temperatures $\omega_c\lesssim T\ll\Delta_0$ one can
disregard the bound states and then one obtains, generally speaking, the higher power
laws for all the components of thermal conductivity.
Since experiments have shown that in several heavy fermion superconductors the
thermal conductivity vanishes with higher powers of $T$ at low temperatures
(see, for example \cite{flouquet}) the theory for thermal conductivity in
unconventional superconductors has been developed and examined in detail
\cite{pet,varma,hir1,mon1,hir2,arfi1,arfi2,arfi3,hir3,hir4,norhir,sauls}.
The experimental
results for $UPt_3$ are in correspondence with the strong scattering from the
impurities, close to the unitarity limit. Some physical discussion of the
strength of the impurity scattering in $UPt_3$ is in \cite{pet,varma}.

Recent experimental results for $UPt_3$ obtained in
\cite{lussier1,lussier2} have drawn much attention\cite{hir3,hir4,norhir,sauls},
as they gave some further useful information about the order parameter
structure in this heavy fermion hexagonal superconductor. In particular, it has
been observed that the anisotropy ratio $\kappa_c/\kappa_b$ of the thermal
conductivity components in $UPt_3$ does not vanish at low temperatures (down to
$T_c/10$). The temperatures, at which one can disregard the influence of
impurity bound states on the transport coefficients, depend upon
the impurity concentration. It has been stressed earlier \cite{arfi2} that for
superconducting $UPt_3$ this influence usually may be neglected just down to
$T_c/10$. It
is confirmed by the absence of a linear term in the temperature dependence of
the thermal conductivity observed in \cite{lussier2} for temperatures down to
$T_c/10$ as well.

Based on this fact below we give a simple qualitative examination of the thermal
conductivity at low temperatures for pure unconventional superconductors
at temperatures $\omega_c\lesssim T\ll\Delta_0$,
separating the contributions from points and lines of nodes (and from the
rest of the Fermi surface, when it is needed) and permitting
the order parameter to have zeroes of different orders at the nodes. Such a
consideration results in a simple condition for the realization of a finite
anisotropy ratio $\kappa_c/\kappa_b$ at low temperatures.
We believe that our simple consideration of the thermal conductivity at low
temperatures $\omega_c\lesssim T\ll\Delta_0$, which permits to discuss
qualitatively a wide number of basis functions for various types of pairing,
complements in some aspects several recent quantitative investigations of this
problem (see, for example,\cite{hir3,norhir,sauls}). On the basis of
self-consistent description of the impurity scattering in unconventional
superconductors, the numerical calculations are performed there for all
temperatures and for several particular basis functions, which are considered
now to be of special interest for the study of superconductivity in $UPt_3$
and in HTSC as well.

So, lets neglect the impurity bound states and consider only essentially
anisotropic superconducting states which meet the condition
$\sum_{{\bf p}}\hat\Delta(\bbox{p})/(E^2-E_{{\bf p}}^2)=0$. Then one has for the
thermal conductivity of unconventional superconductor in the case of resonance
scattering (see e.g. \cite{hir1,arfi2})
\begin{equation}
\frac{\displaystyle\kappa_{ij}}{\displaystyle\kappa_N(T_c)}=
\frac{\displaystyle 18T}{\displaystyle \pi^2T_c}\int_0^{\infty}dE\left(
\frac{\displaystyle E}{\displaystyle T}\right)^2\left[-
\frac{\displaystyle\partial n^0(E)}{\displaystyle \partial E}\right]
\frac{|g(E)|^2}{Re\,g(E)}\int_{|\Delta({\bf p})|\le E}
\frac{\displaystyle d\Omega}{\displaystyle 4\pi}
\frac{\displaystyle p_ip_j}{\displaystyle p^2}
\frac{\displaystyle \sqrt{E^2-|\Delta(\bbox{p})|^2}}{\displaystyle E} .
\label{thc}
\end{equation}
Here ${\kappa_N(T_c)}$ is the thermal conductivity in the normal state at
$T=T_c$, $n^0(E)$ is the equilibrium Fermi distribution function for
excitations, and the function $g(E)$ is defined as follows
\begin{equation}
g(E)=\int\frac{\displaystyle d\Omega}{\displaystyle 4\pi}
\frac{\displaystyle E}{\displaystyle \sqrt{E^2-|\Delta(\bbox{p})|^2}} .
\label{g}
\end{equation}

One can easily see that due to the factor
$\partial n^0(E)/\partial E $ at low temperatures $T\ll \Delta_0$ only
low energy behavior  of the integrand in (\ref{thc}) is of interest
(for $E\lesssim T$). Then only narrow regions near the nodes of the
order parameter are essential for the integral over the momentum directions
in (\ref{thc}) and for the real part of the function $g(E)$. But, generally
speaking, it is not the case for the imaginary part of $g(E)$ at low energy
(see (\ref{g})). Due to this fact and contrary to the case of specific heat, for
the thermal conductivity at low temperatures $T\ll \Delta_0$  contributions
from all the Fermi surface -- not only from the narrow vicinities of nodes of
the order parameter on the Fermi surface -- may be of
importance. Nevertheless the contributions can be divorced from each other and
specificated, and in the case of zeroes of higher orders at the nodes only
those narrow regions become dominating. Note that regions far from nodes, as it
is seen from (\ref{g}), result in the imaginary part of $g(E)$ at low energy
which is simply a linear function of energy.

 Let's consider, for instance, the order parameter, which behaves  near
the line of nodes as $|\Delta(\bbox{p})|=\Delta_0|\theta-\pi/2|^n$ \quad
($|\theta-\pi/2|\ll 1$, $ n>0$) and near the point node as $|\Delta(\bbox{p})|=
\Delta_0\theta^m$ \quad ($\theta\ll 1$, $m>0$). Then one gets at low energy
$Re\,g(E),Im\,g(E)\propto (E/\Delta_0)^{1/n}$,\quad $(E/\Delta_0)^{2/m}$
respectively, but in the case of imaginary part of $g(E)$ this estimate is
valid only if $n>1$, $m>2$. For values
$m<2$ and $n<1$ one obtains for the main term $Im\,g(E)\propto E$  which
results from regions lying far from the nodes of the order parameter on the
Fermi surface. So, both $Re\,g(E)$ and
$Im\,g(E)$ should be taken into account in considering the thermal conductivity
at low temperatures. In the case $n>1$, $m>2$ only the behavior of the order
parameter near nodes turns out to be essential; for $n<1$, $m<2$ contributions
from all the Fermi surface are of importance. For specific and quite important
particular values $n=1$ and $m=2$ one gets apart from the linear terms in $E$
for $Re\,g(E)$, $Im\,g(E)$ also a contribution from the vicinity of nodes
$Im\,g(E)\propto (E/\Delta_0)\ln(E/\Delta_0)$. As in the case of specific heat
a pure logarithmic approximation is a very strong restriction on
the magnitude of energy and usually logarithmic factors must be taken into
consideration together with constants. The anisotropy of the thermal
conduction is defined by the integral over momentum directions in (\ref{thc})
as well. As a result one gets for temperatures
$\omega_c\lesssim T\ll\Delta_0$:
$$
\kappa_{zz}=
L_z\left\{
\begin{array}{cl}
T^{1+4/n},&n>1\\ T^5(\ln^2(T/\Delta_0)+l_{z1}\ln(T/\Delta_0)+
l_{z2}),&n=1\\ T^{2/n+3},& n<1
\end{array}
\right.+\qquad \qquad \qquad \qquad \qquad \qquad \qquad
$$
\begin{equation}
\qquad \qquad \qquad \qquad \qquad \qquad \qquad +P_z\left\{
\begin{array}{cl}
T^{1+4/m},&m>2\\ T^3(\ln^2(T/\Delta_0)+p_{z1}\ln(T/\Delta_0)
+p_{z2}), &m=2\\
T^3,&m<2 ,
\end{array}
\right.
\label{kz}
\end{equation}

$$
\kappa_{xx}=\kappa_{yy}=
L_x\left\{
\begin{array}{cl}
T^{1+2/n},&n>1\\ T^3(\ln^2(T/\Delta_0)+l_{x1}\ln(T/\Delta_0)+
l_{x2}),&n=1\\ T^{3},& n<1
\end{array}
\right.+\qquad \qquad \qquad \qquad \qquad \qquad
$$
\begin{equation}
\qquad \qquad \qquad \qquad \qquad \qquad \qquad +P_x\left\{
\begin{array}{cl}
T^{1+6/m},&m>2\\ T^4(\ln^2(T/\Delta_0)+p_{x1}\ln(T/\Delta_0)
+p_{x2}), &m=2\\
T^{3+2/m},&m<2 .
\end{array}
\right.
\label{kx}
\end{equation}

The coefficients $L$, $l_{i1}$, $l_{i2}$ and  $P$, $p_{i1}$, $p_{i2}$ ($i=x,z$)
correspond here to the contributions from the line of nodes and the point node
respectively.

It follows from Eqs.(\ref{kz}), (\ref{kx}), that under the condition $m=2n$
the main contribution to $\kappa_{zz}$ at low temperatures originates from
the point node, while in the case of $\kappa_{xx}$ the contribution from the
line of nodes dominates. Taking this fact into account one obtains in
qualitative agreement with experimental results the finite (temperature
independent) value for the anisotropy ratio $\kappa_{zz}/\kappa_{xx}$ at low
temperatures under the condition $m=2n$. It is worth noting that in the case
of hexagonal superconductor an important particular example for the type of
pairing with $m=2n$ (and integer $n$, $m$) is a $E_{2u}$-representation (in the
simplest case $m=2n=2$).  At the same time
in the simplest case of $E_{1g}$-representation one should put $n=m=1$
and then it follows from Eqs.(\ref{kz}),(\ref{kx}), that the ratio
$\kappa_{zz}/\kappa_{xx}$ manifests only feebly marked logarithmic temperature
dependence at low temperatures, which possibly can not be determined now
experimentally within the temperature interval $\omega_c\lesssim T\ll\Delta_0$.

From this qualitative discussion one can arrive at the conclusion,
that the both representations $E_{1g}$ and $E_{2u}$ with the simplest particular
basis functions do not contradict to the finite value of the thermal
conductivity anisotropy ratio at temperatures $\omega_c\lesssim T\ll\Delta_0$.
In more complicated cases the anisotropy ratio $\kappa_{zz}/\kappa_{xx}$ in
fact doesn't depend upon temperature within the temperature region considered,
if the indices $n$ and $m$ satisfy the condition $m=2n$. The presence of both
line of nodes and point nodes is essential in order to form the observed
temperature dependence of the thermal conductivity.

\section{ Quasiparticle tunneling at low temperature and voltage}

For isotropic s-wave superconductors a small number of
quasiparticles activated above the gap
at low temperatures $T\ll \Delta_0$ gives rise to the junction
conductance which falls off exponentially with decreasing temperature.
At zero temperature and for the externally applied voltage $V<\Delta_0/e$
for NS junctions (and $V<(\Delta_{0+}+\Delta_{0-})/e$ for SS junctions)
the quasiparticle current across the tunnel junctions takes place only for
superconductors with anisotropic order parameter. In the case of very small
value of voltage ($V\ll \Delta_0$) the current occurs only for the
superconductors with nodes of the order parameter on the Fermi surface. Below
we show that respective I-V curves for these junctions essentially depend on
the behavior of the order parameter in the vicinity of nodes, in particular, on
the multiplicities of the nodes.

If the permanent voltage $V$ is applied to the tunnel junction between two
metals, then one gets the following expression for a dissipative contribution
to the current across the junction  in the lowest order in the junction
transparency $D(\hat{\bbox{p}})$ ($\hat{\bbox{p}}=\bbox{p}_F/|\bbox{p}_F|$;\ \
see e. g.\cite{zai,bgz}):
$$
j_N=e\int_{v_x>0}\biggl(\int\biggl{[}\tanh \biggl{(}\frac{E}{2T}\biggr{)}-
\tanh \biggl{(}\frac{E-eV}{2T}\biggr{)}\biggr{]}
g_+(E-eV,\hat{\bbox{p}}_+)
g_-(E,\hat{\bbox{p}}_-)dE\biggr)
$$
\begin{equation}
\times v_x(\hat{\bbox{p}}_-)
D(\hat{\bbox{p}}_-) \frac{d^2S_-}{(2\pi)^3v_F}.
\label{disscur}
\end{equation}
Here $g_{\pm}(E,\hat{\bbox{p}}_\pm )$ are the normalized  densities of states
(for fixed both energy and the momentum direction) of two metals in the
vicinity of a tunnel barrier, the index $+(-)$
labels the right (left) half space with respect to the boundary plane, and
$v_x$ is the Fermi velocity component  along the normal to the plane interface
$\bbox{n}\parallel Ox$. The integration in (\ref{disscur}) is carried out over
the part of the Fermi surface with $v_x>0$. The relation between the incident
and transmitted Fermi momenta ( that is between $\bbox{p}_-$ and $\bbox{p}_+$ )
is as follows. The components parallel to the specular plane interface are equal
to each other, while the values of their normal components are determined by
the  Fermi surfaces of corresponding metals. Naturally, in the particular case
of identical superconductors with the spherical Fermi surfaces the total
incident and transmitted momenta are equal to each other
$\bbox{p}_-=\bbox{p}_+$.
For an anisotropically paired superconductor the tunneling density of states
is essentially dependent upon the orientation of crystalline axes of the metal
relative to the interface. Below we consider, for simplicity, only those
particular crystal orientations, when the order parameter is not suppressed at
the specularly  reflecting interface as compared to its value in the depth of
pure superconductor. These orientations are determined by the equation
$\Delta(\hat{\bbox{p}}) =\Delta(\check{\bbox{p}})$, where $\hat{\bbox{p}}$ \
$(\check{\bbox{p}})$ denotes the direction of the incident (reflected) electron
momentum (see e.g.\cite{bgz}). Then the following expression for the density of
states is valid
\begin{equation}                      g_{\pm}(E,\hat{\bbox{p}}_\pm )=
\frac{|E |\Theta (|E |-|\Delta _{\pm}(\hat{\bbox{p}}_{\pm})|)}
{\sqrt{E^2-|\Delta_{\pm}(\hat{\bbox{p}}_{\pm})|^2}}.
\label{den}
\end{equation}

Under the condition $T\ll eV$ Eq.(\ref{disscur}) is transformed after
substitution (\ref{den}) into (\ref{disscur}) to the expression
$$
j_N=2e^2V\int_{v_{x}>0} \int_{0}^{1}d\omega\frac{\omega(1-
\omega)\Theta\biggl(\omega-\biggl\vert\Delta_-(\hat{\bbox{p}}_-)/ eV\biggr\vert\biggr)
\Theta\biggl(1-\omega-\biggl\vert\Delta_+(\hat{\bbox{p}}_+)/ eV\biggr\vert\biggr)}{
\sqrt{\omega^2-|\Delta_-(\hat{\bbox{p}}_-)|^2/(eV)^2}
\sqrt{(1-\omega)^2-|\Delta_+(\hat{\bbox{p}}_+)|^2/(eV)^2}}
$$
\begin{equation}
\times D(\hat{\bbox{p}}_-) v_{x}(\hat{\bbox{p}}_-)\frac{d^{2}S_-}{(2\pi)^{3}
v_{F}}.
\label{50}
\end{equation}

One can easily see that for SS junction in the case $|eV|\ll \Delta_{0\pm}$ the
contribution to the integral in (\ref{50}) comes entirely from the narrow
vicinities of the directions to the common nodes of two order parameters, as if
they were drawn on the same Fermi surface in
taking account of the relative orientation of the superconductors. These
directions are defined by the relation $|\Delta_-(\hat{\bbox{p}}_-)|=
|\Delta_+(\hat{\bbox{p}}_+)|=0$. Let the latter condition results in the
momentum direction $\hat{\bbox{p}}_i$ which corresponds to a point
of the intersection on the Fermi surface of two lines
of nodes of the order parameters $\Delta_-$ and $\Delta_+$ respectively. We
suppose that the behavior of the order parameters close to this point may be
represented as follows
\begin{equation}
|\Delta_-(\hat{\bbox{p}})|=\Delta_{0-}|\gamma_-|^n, \ \ \ \
|\Delta_+(\hat{\bbox{p}})|=\Delta_{0+}|\gamma_+|^m, \ \ \ \ n,m>0,
\label{intersec}
\end{equation}
where $\gamma_{\mp}$ are the angles which are counted from
$\hat{\bbox{p}}_i$ in the directions perpendicular to the lines of nodes of the
corresponding order parameters $\Delta_{\mp}$. Substituting (\ref{intersec})
into (\ref{50}) we obtain in the particular case of spherical Fermi surface the
leading order contribution to the quasiparticle current for $eV\ll \Delta_0$ :
\begin{equation}
j_N=a\frac{ep_F^2}{v_F|\sin \chi|}v_{x}(\hat{\bbox{p}}_i) D(\hat{\bbox{p}}_i)
(eV)|eV/\Delta_0|^{\frac{1}{n}+\frac{1}{m}}.
\end{equation}
Here $a$ is a numerical factor and $\chi$ is the angle between two lines of
nodes in the point of their intersection. The angle $\chi$ is supposed to obey
the condition $\chi\gg (eV/\Delta_0)^{1/m}$ ( for $m>n$). In the opposite limit
the lines coincide and then one gets $j_N\propto V^{1+(1/n)}$.
 Summing up the contributions from all intersection points one finds
\begin{equation}
j_N= \frac{\tilde a}{R_N}V|eV/\Delta_0|^{\frac{1}{n}+\frac{1}{m}},
\label{intersecpoints}
\end{equation}
where $R_N$ is the normal-state resistance of a tunnel junction and the
numerical factor $\tilde a$ depends on the relative orientation of
superconductors.

We see that the index of the power law (\ref{intersecpoints}) for the I-V curve
in the case of SS junction
at law voltage and temperature depends essentially upon the multiplicities
of nodes of the order parameters. In the particular case $n=m=1$ the result
(\ref{intersecpoints}) is reduced to that obtained earlier in\cite{bgz}.

The quasiparticle current for the case of NS junction may be obtained by
substituting $\Delta_+=0$ into (\ref{50}) and integrating over $\omega$
\begin{equation}
 j_N=2e^2V  \int_{v_{x}>0}
 \sqrt{1-\frac{\displaystyle |\Delta(\hat{\bbox{p}})|^2}{(eV)^2}}
 \Theta \left(1-\biggl\vert\frac{\Delta (\hat{\bbox{p}})}{eV}\biggr\vert\right)
          D(\hat{\bbox{p}}) v_{x}(\hat{\bbox{p}})\frac{d^{2}S}{(2\pi)^{3} v_{F}}.
\label{44}
\end{equation}
If the externally applied voltage is small enough, so that $|eV|\ll \Delta_0$,
the contribution to $j_N$ in (\ref{44}) comes from narrow vicinity of nodes of
the order parameter. One can easily verify that if the order parameter near the
lines of nodes on the spherical Fermi surface has the form
$|\Delta(\hat{\bbox{p}})|=\Delta_0|\theta-\theta_0|^n$ or
$|\Delta(\hat{\bbox{p}})|=\Delta_0|\varphi-\varphi_0|^n$, it follows from
(\ref{44}) for the quasiparticle current across NS junction
\begin{equation}
j_N\propto e^2V|eV/\Delta_0|^{1/n}.
\label{jNlines}
\end{equation}

Analogously, in the case of the order parameter behavior
$|\Delta(\bbox{p})|=|\Delta_{01}(\varphi-\varphi_0)^m+\Delta_{02}(\theta-
\theta_0)^n|$ close to the point node in the direction $\hat{\bbox{p}}_0$, one
gets
\begin{equation}
j_N\propto \frac{V}{R_N}|eV/\Delta_0|^{\frac{1}{n}+\frac{1}{m}}.
\label{jNpoint}
\end{equation}

At last, for the order parameter behavior near the pole
$|\Delta(\hat{\bbox{p}})|=\Delta_0|\phi-\phi_0|^m\theta^{2n}$  the
contribution to the quasiparticle current is given by
\begin{equation}
j_N\propto e^2D(\hat{\bbox{p}}_{pole}) v_{x}(\hat{\bbox{p}}_{pole})
\left\{
\begin{array}{ll}
V\left\vert\frac{\displaystyle eV}{\displaystyle \Delta_0}\right\vert^{1/m},\ \ \ &m>n,\\
\\
V\left\vert\frac{\displaystyle eV}{\displaystyle \Delta_0}\right\vert^{1/m}
\left(\ln \left\vert\frac{\displaystyle\Delta_0}{\displaystyle eV}\right\vert+
c\right),\ \ \
&m=n,\\
\\
V\left\vert\frac{\displaystyle eV}{\displaystyle \Delta_0}\right\vert^{1/n},\ \
\ &m<n.\\
\end{array}
\right.
\label{pole}
\end{equation}

The results (\ref{jNlines}) and (\ref{pole}) in the particular cases $n=1$ and
$m=n=1$ are reduced to those obtained earlier in\cite{bgz}.

\section{Anisotropy of the boundary conditions
for a superconducting order parameter}

Many aspects of the problem of boundary conditions for the order parameter
of anisotropically paired superconductors have already been studied in the
literature (see, e.g.\cite{ambderai,kurki,samokh,bgz,bgs} and references
therein). Below we discuss one characteristic feature of the boundary
conditions near $T_c$ , which has been recently noticed in \cite{bgz} for the
case of specularly reflecting boundary. For simplicity, let's consider a singlet
pairing and a one component anisotropic order parameter in the vicinity of the
plane specular and fully reflecting boundary at $x=0$. For any crystal
orientation with respect to the interface the boundary condition may be written
in the form $q\eta'(0)=\eta(0)$. It turns out that with the aid of general
symmetry arguments and, for example, irrespective of the shape of the
Fermi surface of the given symmetry, one can unambiguously fix only those
crystal orientations for which the parameter $q$ is equal to zero and infinity.
The value $q=\infty$ corresponds to the boundary condition $\eta'(0)=0$
coinciding with the condition for an isotropic s-wave superconductor
at an impenetrable boundary, where the relation $\eta(0)=\eta_\infty$ is valid.
Here $\eta_\infty$ is the value of the superconducting order parameter
far from the boundary. For an anisotropically paired
superconductor one has $q=\infty$
only for those orientations of the boundary for which the relation
$\Delta(\hat{\bbox{p}})=\Delta(\check{\bbox{p}})$ holds. The value $q=0$
is realized under the condition
$\Delta(\hat{\bbox{p}})=-\Delta(\check{\bbox{p}})$, in which case the boundary
condition takes the form $\eta(0)=0$ and the order parameter is fully suppressed
at the interface. As earlier $\hat{\bbox{p}}$ \ $(\check{\bbox{p}})$ denotes here
the direction of the incident (reflected) electron momentum.
For the point symmetry group $D_{4h}$ of the crystal there are
five planes of symmetry, and the order parameter transforms according to one of
the irreducible representations of this group. So, if the normal to the boundary
is perpendicular to the symmetry plane and the character of the irreducible
representation is 1 (-1), the boundary condition is $\eta'(0)=0$ ($\eta(0)=0$).

For other orientations the detailed form of $q$
as a function of crystal orientation relative to the boundary plane for the
given type of pairing essentially depends on the particular form of
the basis function $\Delta(\hat{\bbox{p}})$ in the depth of superconductor.
Since the characteristic value for the parameter $q$ is $\xi(T)$ one can divide
all orientations into two regions where $q<\xi(T)$ and $q>\xi(T)$
respectively. At first sight the characteristic scales for these two regions
could be comparable on the order of values. However, it is not the case near
$T_c$ and may be valid only for low temperatures. E.g. for the particular basis
function $p_{x0}^2-p_{y0}^2$ ( pertaining obviously to the pairing symmetry
of the
type $p_{x0}^2-p_{y0}^2$) and the spherical Fermi surface, the values $\eta(0)$
and $\eta_\infty$  are found to be of the same order of magnitude (and hence
$\xi(T)\lesssim q$) only for particular orientations  of the normal to the
boundary $\bbox{ n}$ within narrow angular intervals
$\Delta\phi_0\sim \Delta\theta_0\sim (\xi_0/\xi(T))^{1/2}$
around the crystal axes $x_0, y_0, z_0$. For other crystal orientations one has
$\eta(0)\sim\eta_\infty\xi_0/\xi(T)\ll \eta_\infty$. Thus, near $T_c$ the d-wave
superconducting order parameter proves to be strongly suppressed in
the vicinity of the insulating barrier, except for the small part of
orientations within the narrow angular intervals mentioned above. From this
point of view the qualitative difference between the boundary conditions for
specular and diffusive interfaces near $T_c$ manifests only within those narrow
angular intervals. Indeed, in the case of diffusive scattering at the interface
one has $q\sim \xi_{0}$ irrespective of the orientations\cite{sha} and, hence,
near $T_c$ the parameter
$\eta (0)\sim\xi_{0}\eta_{\infty}/\xi (T)\ll \eta_{\infty}$ for all crystal
orientations. So, near $T_c$ narrow peaks $\xi_0\ll\xi(T)\lesssim q$ of the
parameter $q$, which occur in the case of specularly reflected boundaries, are
simply cut if the boundary is diffusive.

It is of interest to consider whether the choice of a different basis function
pertaining to the same symmetry type of pairing can result near $T_c$ in the
expansion of the angular regions where the parameter $q$ has large values
$\xi(T)\lesssim q$. Below we show that it is not possible for quite a
wide set of the bases functions. In particular, for the bases functions with
higher multiplicities of the nodes on the Fermi surface the angular regions
with $\xi(T)\lesssim q$ prove to be more narrow, not wider.

Microscopic consideration based, in particular, on some variational procedure,
leads to the following expression for the parameter $q$ near
$T_c$\cite{bgz,samokh,bgs}:
$$
q= \xi_0\left ( \frac{\pi^{3}}{336\zeta(3)}
\int_{v_{x}>0} |\psi(\hat{\bbox{p}})+\psi(\check{\bbox{p}})|^{2}\hat{v}_{x}^{3}
d^{2}S+
\frac{7\zeta(3)}{4\pi^{3}}
\frac{(
\int_{v_{x}>0} |\psi(\hat{\bbox{p}})+\psi(\check{\bbox{p}})|^{2}\hat{v}_{x}^{2}
d^{2}S
)^{2}}{
\int_{v_{x}>0} |\psi(\hat{\bbox{p}})-\psi(\check{\bbox{p}})|^{2}\hat{v}_{x}
d^{2}S
}\right )
$$
\begin{equation}
\times\left ( \int|\psi(\hat{\bbox{p}})|^{2}\hat{v}_{x}^{2}d^{2}S \right )^{-1}.
\label{q}
\end{equation}
Here integration  over the Fermi surface is confined to its part with $v_x>0$
and the following notations are introduced:\ $\hat{v}_x= v_{Fx}/v_F$,\
$\xi_0=v_F/T_c$. The superconducting order parameter for the inhomogeneous state
near $T_c$ is supposed to be represented in the conventional form
$\Delta(\hat{\bbox{p}},\bbox{r})=\psi(\hat{\bbox{p}})\eta(\bbox{r})$.

For isotropic superconductors the basis function $\psi$ doesn't depend upon the
momentum at all and one gets from (\ref{q}) the equality $q=\infty$. For
anisotropically paired superconductors there are, as a rule, only several
isolated and governed by the symmetry crystal orientations relative to the
boundary for which
$q=\infty$ and, hence, $\psi(\hat{\bbox{p}})=\psi(\check{\bbox{p}})$ for all
$\hat{\bbox{p}}$. The crystal orientation with respect to the boundary may be
characterized by the unit vector $\bbox{n}$ which describes the orientation of
the normal to the boundary relative to the crystalline axes $x_0, y_0, z_0$.
Denoting by $\bbox{n}_0$ the orientation corresponding to the value $q=\infty$,
we are interested now in the characteristic scale of the angular regions around
the direction $\bbox{n}_0$ where  $\xi(T)\lesssim q$, i.e. the parameter $q$
has quite a large value yet.

It follows from Eq.(\ref{q}), that the relation $\xi(T)\lesssim q$ is valid
under the condition
\begin{equation}
\frac{\displaystyle \xi(T)}{\displaystyle \xi_0}\lesssim\frac{
\int_{v_{x}>0} |\psi(\hat{\bbox{p}})+\psi(\check{\bbox{p}})|^{2}\hat{v}_{x}^{2}
d^{2}S
)}{
\int_{v_{x}>0} |\psi(\hat{\bbox{p}})-\psi(\check{\bbox{p}})|^{2}\hat{v}_{x}
d^{2}S } \ .
\label{con}
\end{equation}

Since in the case $\bbox{n}=\bbox{n}_0$ one gets $\psi(\hat{\bbox{p}})=
\psi(\check{\bbox{p}})$ (in the particular case when the boundary coincides
with the symmetry plane of the Fermi surface, one has
$\check{\bbox{p}}=\hat{\bbox{p}}-2(\hat{\bbox{p}}\bbox{n})\bbox{n}$),
in the narrow vicinity of this direction
one obtains $\psi(\check{\bbox{p}})\approx\psi(\hat{\bbox{p}})+
(\delta\bbox{n}\bbox{\nabla}_{{\bf n}})\psi(\check{\bbox{p}})
\vert_{{\bf n}={\bf n}_0}$. Here
$\delta\bbox{n}=\bbox{n}-\bbox{n}_0$ and one has
$\widehat{(\bbox{n},\bbox{n}_0)}\approx|\delta\bbox{n}|$ for the angle between
vectors $\bbox{n}$ and $\bbox{n}_0$. Substituting these relations into
(\ref{con}), we obtain the following restriction for the angle
$\widehat{(\bbox{n}, \bbox{n}_0)}$:
\begin{equation}
\widehat{(\bbox{n}, \bbox{n}_0)}\lesssim \left(
\frac{\displaystyle \xi_0}{\displaystyle \xi(T)}\right)^{1/2}
\frac{(
\int_{v_{x}>0} |\psi(\hat{\bbox{p}})|^{2}\hat{v}_{x}^{2}
d^{2}S
)^{1/2}}{\left(
\int_{v_{x}>0} |(\bbox{e}_{\delta\bbox{n}}\bbox{\nabla}_{{\bf n}})
\psi(\check{\bbox{p}})\vert_{{\bf n}={\bf n}_0}|^{2}\hat{v}_{x}
d^{2}S
\right )^{1/2}} \ ,
\label{est}
\end{equation}
where $\bbox{e}_{\delta\bbox{n}}$ is the unit vector along the direction
$\delta\bbox{n}$.

We see that the appearance of the small parameter $(\xi_0/\xi(T))^{1/2}$ is a
quite general characteristic feature for the estimation of the angle
$\widehat{(\bbox{n}, \bbox{n}_0)}$, which is independent upon the particular
form
of the basis functions. In order to estimate the other factor by which the first
one is multiplied in (\ref{est}), let's consider the following set  of the
basis functions for the tetragonal unconventional superconductors
\begin{equation}
\psi(\hat{\bbox{p}})=\Delta_0\tanh\left\{ \mu(\hat{p}_{x0}^2-
\hat{p}_{y0}^2)^n\right\} \ , \ \ \ \ \ n=1,2,\dots \ ,
\label{set}
\end{equation}
where $\mu$ and $n$ are the parameters. All functions with odd values of
$n$ belong evidently to the same type of pairing as the function
$(\hat{p}_{x0}^2-\hat{p}_{y0}^2)$. Analogously, the functions with even values
of $n$ pertain to the unit representation of the point tetragonal group
$D_{4h}$ (then respective nodes of the order parameter, of course, are
accidental and not associated with the symmetry type of pairing ).

The substitution of (\ref{set}) into (\ref{est}) results in the following
estimation for the angular intervals around $\bbox{n}_0$, where
$\xi(T)\lesssim q$:
\begin{equation}
\Delta\varphi_0, \Delta\theta_0\sim
\frac{\displaystyle 1}{\displaystyle \nu n}\left(
\frac{\displaystyle \xi_0}{\displaystyle \xi(T)}\right)^{1/2} \ ,
\end{equation}
where $\nu=\max(1,\mu)$.

It follows from here that the angular intervals can't be essentially more
than $(\xi_0/\xi(T))^{1/2}$, but may be more narrow, for instance, in the case
of higher multiplicities of the nodes of the order parameter.

\section{Conclusions}

Density of states at low energy, the specific heat in homogeneous and mixed
states at low temperatures, the low temperature behavior of the thermal
conductivity, the penetration depth  and I-V curves for the quasiparticle
current at low voltage have been examined for pure anisotropically paired
superconductors having various multiplicities of the nodes of the order
parameter.
A specific anisotropy of the boundary conditions for
unconventional superconducting order parameter near $T_c$ for the case of
specular reflection from the boundary was also investigated.
Since the multiplicites may be different within the same symmetry
type of pairing,
the problem of an unambiguous identification of a type of superconducting
pairing on the basis of corresponding experimental results
becomes more complicated and should be considered taking into account
the results obtained above.

\section*{Acknowledgements}
One of us (Yu.S.B.) wishes to thank J.A. Sauls and S. Yip
for stimulating discussions. This work was supported by the grant
No. 94-02-05306
of the Russian Foundation for Basic Research. A.A.S. acknowledges
Forschungszentrum J\"ulich and International Soros Science Education
Program for financial support.


\begin{references}
\bibitem{lussier1} B. Lussier, B. Ellman, and L. Taillefer, Phys. Rev. Lett.
{\bf 73}, 3294 (1994).
\bibitem{lussier2} B. Lussier, B. Ellman, and L. Taillefer, preprint
( cond-mat/9504072 ) (1995).
\bibitem{uedarice} K. Ueda and T.M. Rice, in {\it Theory of Heavy Fermions and
Valence Fluctuations}, edited by T. Kasuya and T. Saso (Springer,Berlin,1985).
\bibitem{hir2} P. Hirschfeld, P. W\"olfle , and D. Einzel, Phys. Rev. B
{\bf 37}, 83 (1988).
\bibitem{muz1} G. Preosti, H. Kim, and P. Muzikar, Phys.Rev. B {\bf 50}, 1259
(1994).
\bibitem{saulyip} S.K. Yip, and J.A. Sauls, Phys. Rev. Lett. {\bf 69}, 2264
(1992).
\bibitem{vol} G.E. Volovik, JETP Lett. {\bf 58}, 469 (1993) (Pis'ma Zh. Eksp.
Teor. Fiz. {\bf 58}, 457 (1993)).
\bibitem{mol} K.A. Moler, D.J. Baar, J.S. Urbach, R. Liang, W.N. Hardy, and
A. Kapitulnik, Phys. Rev. Lett. {\bf 73}, 2744 (1994).
\bibitem{mol2} K.A. Moler, A. Kapitulnik, D.J. Baar, R. Liang, W.N. Hardy,
preprint (cond-mat/9505129) (1995).
\bibitem{fetter} A.L. Fetter, and P.C. Hohenberg, in {\it Superconductivity},
edited by R.D. Parks (M. Dekker, New York, 1969).
\bibitem{hhess}H.F. Hess, R.B. Robinson, and J.V. Waszczak, Phys. Rev. Lett.
{\bf 64}, 2711 (1989).
\bibitem{fischer} Ch. Renner, and \O. Fischer, Phys. Rev. B {\bf 51}, 9208
(1995).
\bibitem{rem1} See, for example,\cite{maki}, where the anisotropy and
inhomogeneity of the density of states at the distances $r\lesssim\xi_0$ has
been considered (on the basis of the Eilenberger equations, some variational
procedure and numerical calculations) for the simplest case of
$(p_{x}^2-p_{y}^2)$-pairing and only the particular value $n=1$.
\bibitem{maki} N. Schopohl, and K. Maki Phys. Rev. B {\bf 52}, 490, (1995).
\bibitem{gross} F. Gross , B.S. Chandrasekhar, D. Einzel, P.J. Hirshfeld,
K. Andres, H.R. Ott, Z. Fisk, J. Smith, and J. Beuers, Z. Phys. B {\bf 64},
175 (1986).
\bibitem{choimuz} C. Choi, and P. Muzikar, Phys. Rev. B {\bf 37}, 5947 (1988).
\bibitem{procar} M. Prohammer, and J.P. Carbotte, Phys. Rev. B {\bf 43}, 5370
(1991).
\bibitem{hirgol} P. Hirschfeld, and N. Goldenfeld, Phys. Rev. B {\bf 48}, 4219
(1993).
\bibitem{xu} D. Xu, S.K. Yip, and J.A. Sauls, Phys. Rev. B {\bf 51}, 16233
(1995).
\bibitem{flouquet} A. Sulpice, P. Gandit, J. Chaussy, J. Flouquet, D. Jaccard,
P. Lejay, and J.L. Tholence, J.Low Temp. Phys. {\bf 62}, 39 (1986).
\bibitem{pet} C.J. Pethick and D. Pines , Phys. Rev. Lett. {\bf 57}, 118 (1986).
\bibitem{varma} S. Schmitt-Rink, K. Miyake, and C.M. Varma , Phys. Rev. Lett.
{\bf 57}, 2575 (1986).
\bibitem{hir1} P. Hirschfeld, D. Vollhardt, and P. W\"olfle , Solid State
Commun. {\bf 59}, 111 (1986).
\bibitem{mon1} H. Monien, K. Scharnberg, L. Tewordt, and D. Walker,
Solid State Commun. {\bf 61}, 581 (1987).
\bibitem{arfi1} B. Arfi, H. Bahlouli, C.J. Pethick, and D. Pines , Phys. Rev.
Lett. {\bf 60}, 2206 (1988)
\bibitem{arfi2} B. Arfi, C.J. Pethick , Phys. Rev. B {\bf 38}, 2312 (1988).
\bibitem{arfi3} B. Arfi, H. Bahlouli, and C.J. Pethick , Phys. Rev. B {\bf 39},
8959 (1989).
\bibitem{hir3} A. Fledderjohann and P.J. Hirschfeld , Solid State Commun.
{\bf 94}, 163 (1995).
\bibitem{hir4} L.S. Borkowski, P.J. Hirschfeld and W.O. Putikka , Phys. Rev. B
{\bf 52}, R3856 (1995).
\bibitem{norhir} M.R. Norman, and P.J. Hirschfeld , preprint
( cond-mat 9509045) (1995).
\bibitem{sauls} M.J. Graf, S.-K. Yip, J.A. Sauls, and D. Rainer, preprint
( cond-mat/9509046 ) (1995).
\bibitem{zai}A.V. Zaitsev, Zh. Eksp. Teor. Fiz. {\bf 86}, 1742 (1984) [Sov.
Phys. JETP {\bf 59}, 863 (1984)].
\bibitem{bgz}Yu.S. Barash, A.V. Galaktionov, and A.D. Zaikin, Phys. Rev. B
{\bf 52}, 665 (1995).
\bibitem{ambderai} V. Ambegaokar, P.G. de Gennes, and D. Rainer, Phys. Rev. A
{\bf 9}, 2676 (1974).
\bibitem{kurki}J. Kurkijarvi, D. Rainer, and J.A. Sauls, Can. J. Phys. {\bf 65},
1440 (1987).
\bibitem{samokh} K.V. Samokhin, Zh. Eksp. Teor. Fiz. {\bf 107}, 906 (1995) [Sov.
Phys. JETP {\bf 80}, 515 (1995)].
\bibitem{bgs} Yu.S. Barash, A.V. Galaktionov, and A.A. Svidzinsky, Phys. Rev. B
{\bf 52}, 10344 (1995).
\bibitem{sha} E.A. Shapoval, Zh. Eksp. Teor. Fiz. {\bf 88}, 1073 (1985) [Sov.
Phys. JETP {\bf 61}, 630 (1985)].
\end{references}
\end{document}